\documentstyle[prl,aps,epsf,twocolumn]{revtex}
\tighten

\newcommand{\smeq}{\! \! = \!}

\newcommand{\aeff}{a_{\rm eff}}

\newcommand{\Gpeak}{G_{\rm peak}}
\renewcommand{\refname}{}
\newcommand{\biblabel}[1]{[#1]} %
\renewcommand{\references}{%
\ifpreprintsty
\vspace*{-0.1 truein}
\hbox to\hsize{\hss\large \refname\hss}%
\else
\vskip3pt
\hrule width\hsize\relax
\vskip -0.2in
\fi
\list{\biblabel{\arabic{enumiv}}}%
{\labelwidth\WidestRefLabelThusFar  \labelsep4pt %
\leftmargin\labelwidth %
\advance\leftmargin\labelsep %
\ifdim\baselinestretch pt>1 pt %
\parsep  4pt\relax %
\else %
\parsep  0pt\relax %
\fi
\itemsep\parsep %
\usecounter{enumiv}%
\def\theenumiv{\arabic{enumiv}}%
}%
\let\newblock\relax %
\sloppy\clubpenalty4000\widowpenalty4000
\sfcode`\.=1000\relax
\ifpreprintsty\else\small\fi
}

\begin{document}
\draft

\title{
Chaos in Quantum Dots: Dynamical Modulation of Coulomb Blockade Peak Heights
}

\author{Evgenii E. Narimanov,$^{1}$ Nicholas R. Cerruti,$^{2}$ 
Harold U. Baranger,$^{1}$ Steven Tomsovic$^{2}$ }

\address{$^{1}$ Bell Laboratories-- Lucent Technologies,
700 Mountain Ave., Murray Hill NJ 07974}
\address{$^{2}$ Washington State University,
Department of Physics, Pullman WA 99164-2814
}

\date{ \today}

\maketitle

\begin{abstract}
We develop a semiclassical theory of Coulomb blockade peak heights in
quantum dots and show that the dynamics in the dot leads to a large
modulation of the peak height. The corrections to the standard statistical
theory of peak height distributions, power spectra, and correlation
functions are non-universal and can be expressed in terms of the classical
periodic orbits of the dot that are well coupled to the leads. The
resulting correlation function oscillates as a function of peak number in a
way defined by such orbits; in addition, the correlation of adjacent
conductance peaks is enhanced. Both of these effects are in agreement with
recent experiments.
\end{abstract}
\vspace*{-0.05 truein}
\vspace*{-0.15 truein}

The electrostatic energy of an additional electron on a quantum dot-- a
mesoscopic island of confined charge with quantized states-- blocks the flow
of current through the dot, an effect known as the Coulomb
blockade\cite{nato-book}. Current can flow only if two different charge
states of the quantum dot are tuned to have the same energy; this produces a
peak in the the conductance of the dot whose magnitude is directly related
to the magnitude of the wavefunction near the contacts to the dot. Since
dots are generally irregular in shape, the dynamics of the electrons is
chaotic, and the characteristics of Coulomb blockade peaks reflect those of
wavefunctions in chaotic systems\cite{jalabert,nato-book-2,Stopa98}.
Previously, a statistical theory for the peaks was
derived\cite{jalabert,nato-book-2} by assuming these wavefunctions to be
completely random and uncorrelated with each other. The experimental
data\cite{Chang96,MarcusFolk96} for the distribution of the Coulomb blockade
peak heights were found to be in excellent agreement with the predictions of
the statistical theory, thus giving a solid foundation to the conjecture of
effective ``randomness'' of the quantum dot wavefunctions.

It therefore came as a surprise when several recent
experiments\cite{MarcusFolk96,MarcusCronenwett97,MarcusPatel98} demonstrated
large correlations between the heights of adjacent peaks. The effect of
nonzero temperature (when several resonances contribute to the same peak)
was found to be insufficient to account for these
correlations\cite{Alhassid98}. To explain the correlations, the enhancement
due to spin-paired levels\cite{MarcusPatel98,Alhassid98}, due to a decrease
of the effective level spacing found in density functional
calculations\cite{Stopa96}, and due to of level anticrossings in interacting
many-particle systems\cite{Hackenbroich97} were proposed. However, we show
here that peak height correlations arise already within an effective
single-particle picture of the electrons in the quantum dot. The specific
internal dynamics of the dot, even though it is chaotic, modulates the
peaks: because all systems have short-time features, chaos is not equivalent
to randomness.  The predicted dynamical modulation is exactly of the type
needed to explain the
experiments\cite{MarcusFolk96,MarcusCronenwett97,MarcusPatel98}.

To study the non-universal effects of the dynamics of a particular dot, we
derive a relation between the quantum conductance peak height and the
classical periodic orbits in the dot. The main effect is that as a system
parameter varies-- the magnetic field or the number of electrons in the dot
in response to a gate voltage, for instance-- the interference around each
periodic orbit oscillates between being destructive and constructive. When
the interference is constructive for those periodic orbits which come close
to the leads used to contact the dot, the wavefunction is enhanced near the
leads, the dot-lead coupling is stronger, and so the conductance is larger.
Likewise, destructive interference produces a smaller conductance. The
resulting modulation can be substantial, as shown in Fig. 1. Similar
short-time dynamical effects have been noted in other contexts such as
atomic and molecular spectra\cite{gutzwiller,BeimsDelos98,heller}, 
eigenfunction
scarring\cite{heller,KaplanHeller}, magnetotransport in antidot
lattices\cite{antidots},  and tunneling into quantum
wells\cite{e2n_prl,monteiro,e2n_physicad}. Such modulation is completely
omitted in theories in which the wavefunction is assumed to change randomly
as the system changes\cite{jalabert,nato-book-2}.

\begin{figure}[b]
\begin{center}
\leavevmode
\epsfxsize = 8.6cm
\epsfbox{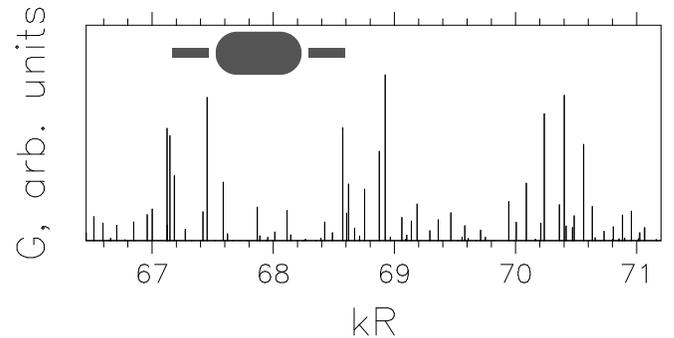}
\end{center}\caption{
The peak conductance from tunneling through subsequent energy levels
for the stadium billiard shown in the inset. Each peak is placed at 
the wavevector $k$ corresponding to its level; $R$ is the radius of
the half-circle parts of the stadium. A Gaussian lead
wavefunction appropriate for tunneling from a single transverse mode
is used with width $ka_{\rm eff} = 15$. 
 }
\end{figure}

Our starting point is the connection between the peak height and the width
of the level in the quantum dot. We consider a dot close to two leads (Fig.
1 inset) so that the width, $\Gamma$, of a level comes from tunneling of the
electron to either lead. When the mean separation of levels is larger than
the temperature $T$ which itself is much larger than the mean width, the
electrons pass through a single quantized level in the dot. The conductance
peak height in this regime is\cite{beenakker}
\begin{eqnarray}
G_{\rm peak} & = & \frac{e^2}{h} \frac{\pi {\Gamma}}{4 kT}.
\label{g_max}
\end{eqnarray}
Here for simplicity we consider symmetrically placed leads-- the total width
is equally split between tunneling to the right and left leads-- spinless
particles, and temperature much smaller than the level spacing.

The width of the level is related by Fermi's Golden Rule to the
square of the matrix element for tunneling between the lead and the dot,
$M^{\ell\rightarrow d}$. A convenient expression for the matrix element in
terms of the lead and dot wavefunctions, $\Psi_\ell$ and $\Psi_d$
respectively, was derived by Bardeen\cite{bardeen} 
and can be expressed as\cite{e2n_prl,monteiro}
\begin{eqnarray}
M^{\ell \rightarrow d} = \frac{\hbar^2}{m^*}  \int_{S} d{\bf S}
 \Psi_{\ell}(r) {\bf \bigtriangledown} \Psi_{d}(r)
\label{me_approx}
\end{eqnarray}
where the surface $S$ is the edge of the quantum dot. $\Gamma$, then,
depends on the square of the normal derivative of the dot wavefunction at
the edge weighted by the lead wavefunction. Writing the product of the
two $\Psi_d$'s in $\Gamma$ as a Green function $G(r,r')$ and using the 
standard
semiclassical expression for the latter\cite{gutzwiller}, we express 
$\Gamma$ as a sum over the classical trajectories which start at $r$ on 
the edge of the dot near the lead and end at $r'$ which is also on the
edge near the lead.

Tunneling from the lead to the dot is dominated by the lowest transverse
energy subband in the constriction between the lead and the
dot\cite{nato-book-2}. Therefore, for the calculation of the tunneling
matrix element the transverse potential in the tunneling region can be taken
quadratic:  $U_{\ell} \sim \kappa \left(y-y_\ell\right)^2$.  In this case
the transverse dependence of the lead wavefunction is simply a harmonic
oscillator wavefunction, so that at the edge of the dot $\Psi_\ell \sim
c_\ell \exp [ - (y - y_\ell )^2/2 a_{\rm eff}^2 ]$, where $\hat{\bf y}$
represents the direction tangential to the boundary of the dot, $y_\ell$
is the center of the lead and constriction, and the effective width is
$a_{\rm eff} = \sqrt{\hbar}/\sqrt[4]{2 \kappa m^*}$.  While the exact form
of the lead wavefunction is not crucial, the $\hbar$-dependence of the
width is important for the semiclassical argument which follows; note that
$a_{\rm eff} \sim \sqrt{\hbar}$ does not depend on a particular transverse
potential.

Using this information about $\Psi_\ell$ in the expression for $M^{\ell
\rightarrow d}$, we see that the lead wavefunction restricts the integration
to a semiclassically narrow region of width $a_{\rm eff} \sim \sqrt{\hbar}$.
This allows one to express the contribution of the open trajectories
entering the Green function in terms of an expansion near their closed
neighbors. In the resulting expression for $\Gamma$, the contribution of
each of these closed orbits is suppressed by a factor exponentially small in
$\Delta p_y^2$, where $\Delta p_y$ is the change of transverse momentum
after one traversal. This suppression is the effect of the mismatch of the
closed orbit (momentum) with the distribution of transverse momentum at the
lead, which is centered at zero with width $\delta p_\ell \sim \hbar/a_{\rm
eff} \sim \sqrt{\hbar}$ for the lowest subband. Therefore, only closed
orbits with {\it semiclassically} small momentum change $\Delta p$
contribute to the width. This in turn implies that the closed orbit is
located semiclassically close (within a distance $\sim \sqrt{\hbar}$) to a
{\it periodic orbit} for which $\Delta p \equiv 0$. Thus, one can express
the tunneling width in terms of the properties of these periodic orbits,
obtaining\cite{subbands}

\begin{eqnarray}
\Gamma = \bar{\Gamma} +  \sum_{\mu:{\rm p.o.}} A_\mu 
\cos\left( \frac{S_\mu}{\hbar} + \phi_\mu \right) 
\label{gamma_sc}
\end{eqnarray}
where the monotonic part is
\begin{eqnarray}
\bar{\Gamma} = \case{ \sqrt{\pi}}{2} c_\ell^2 a_{\rm eff}
\frac{p^2}{m^*} \
e^{-\zeta} \
\left[I_0\left(\zeta\right)
+ I_1\left(\zeta\right)
\right], \
\ \zeta = \frac{p^2 a_{\rm eff}^2}{2 \hbar^2} ,
\nonumber
\end{eqnarray}
the amplitude is
\begin{eqnarray}
A_\mu & = & 4 \sqrt{2} \ \frac{\hbar c_\ell^2 p^\mu_z}{m^*}
\ \left[{\rm Tr}^2\left[M_\mu\right]
\left(1 + \sigma_+^2 \right)
\left( 1 + \sigma_-^2 \right) \right]^{-1/4}
\nonumber \\
& \times & 
\exp
\left( 
- 
\frac{\sigma_+^2 \bar{p}^2}
     {\left(1 + \sigma_+^2\right)} 
- 
\frac{\sigma_-^2 \bar{y}^2}
     {\left(1 + \sigma_-^2\right) } 
\right)
\nonumber
\end{eqnarray}
with
\begin{eqnarray}
\sigma_\pm  & \equiv & \case{1}{2} \left[
\overline{m}_{12} \! - \! \overline{m}_{21} \pm   
\sqrt{\left(\overline{m}_{22}\! -\! \overline{m}_{11}\right)^2
+ \left(\overline{m}_{21}\! +\! \overline{m}_{12}\right)^2} \right]
\nonumber 
\\
\overline{m}_{ij} & \equiv & 
\frac{2 m^\mu_{ij}}
{ {\rm Tr}\left[M_\mu\right] + 2}
\left( \frac{a_{\rm eff}^2}{\hbar} \right)^{\frac{j-i}{2}}
\nonumber \\
\theta & \equiv & \case{1}{2} \arctan\left(
\frac{\overline{m}_{22} - \overline{m}_{11}}
{\overline{m}_{21} + \overline{m}_{12}}
\right)
\nonumber \\
\bar{y} & \equiv &
\cos\theta \left(y_\mu - y_\ell\right)/a_{\rm eff}
+ \sin\theta \ p^\mu_y a_{\rm eff}/\hbar
\nonumber \\
\bar{p} & \equiv  &
\cos\theta \ p^\mu_y a_{\rm eff}/\hbar
- \sin\theta \left(y_\mu - y_\ell\right)/a_{\rm eff} \, ,
\nonumber 
\end{eqnarray}
and, finally, the result for the slowly varying phase $\phi_\mu$ will be
given elsewhere. Here $I_n$ is the Bessel function of complex argument,
${\bf p}^\mu$ is the electron momentum for the periodic orbit $\mu$ at the
bounce point, $y_\mu$ is the bounce point coordinate, $S_\mu$ is the action
of the periodic orbit, and $M_\mu \equiv \left(m_{ij}^\mu\right)$ is the
corresponding monodromy matrix\cite{gutzwiller}. The semiclassical approach
used here is similar to the calculation of the tunneling current in a
resonant tunneling diode in a magnetic field developed in Ref.
\onlinecite{e2n_prl} and \onlinecite{e2n_physicad}.

The equation above is the main result of this paper: it expresses the
modulation of the heights of the Coulomb blockade peaks by the classical
periodic orbits. Note that the result (\ref{gamma_sc}) is valid not only
for chaotic but also for both integrable and mixed systems (for an
integrable system or for the contributions of the remaining unbroken tori of
a mixed system ${\rm Tr}\left[M\right] \equiv 2$).

In order to assess the validity of the semiclassical expression
(\ref{gamma_sc}), we compare it to numerical calculations for two simple
billiards, one integrable-- the circle-- and one chaotic-- the stadium. The
stadium billiard (Fig. 1 inset) is one of the canonical examples of a
completely chaotic system\cite{gutzwiller}. From the dot wavefunctions
(numerically obtained using the method of Ref. \onlinecite{Julie}), $\Gamma$
can be calculated from Eq. (\ref{me_approx}) using the $\Psi_\ell$ given
above. To observe the variation in peak height, we vary the energy, or
equivalently the wavevector $k \smeq p/\hbar$, which changes the number of
electrons on the dot as more levels are filled.

\begin{figure}[t]
\begin{center}
\leavevmode
\epsfxsize = 8.6cm
\epsfbox{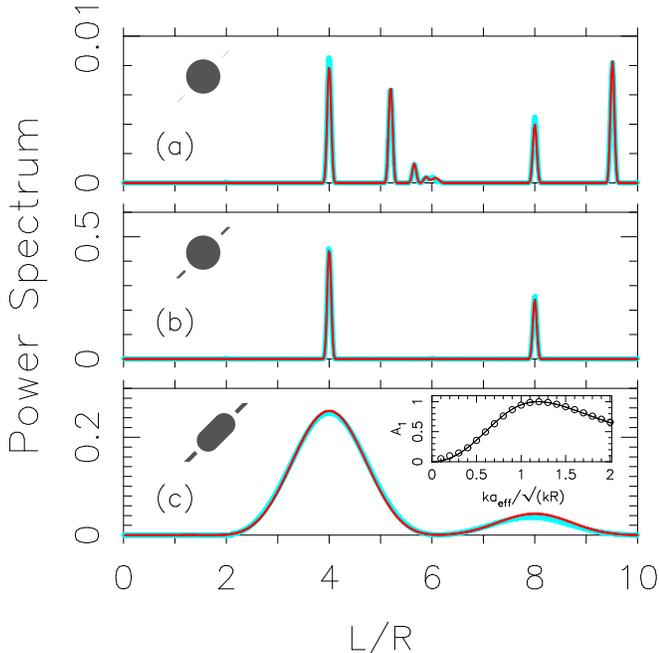}
\protect\caption{
Length spectrum of the oscillations in $G(k)$ obtained from the Fourier
power, numerical (blue) and semiclassical (red) results compared. The power 
is normalized to the mean conductance and then this mean is removed for 
clarity. (a) Circular dot with narrow leads, $ka_{\rm eff}
\approx 1.2$.
where $\aeff$ is the width of the lead wavefunction.
(b) Circular dot with wider leads, $ka_{\rm eff} \approx 12$.
(c) Stadium dot using data in Fig. 1; dependence of amplitude at
$L/R \smeq 4$ on $ka_{\rm eff}$ in inset.
The width of the peaks reflects the length of $G(k)$ used. More data was
available for the circle because it is integrable; conservation of angular 
momentum allows a simple representation of the wavefunctions in terms of 
Bessel functions. In (a) the peak at
$L/R \smeq 4$ is the diameter, that at $8$ is its repetition, those
accumulating to $2\pi$ are the whispering gallery trajectories, and the 
largest length peak is the star orbit.
The magnitude of the oscillatory part compared to the mean depends
on the strength of the coupling to the periodic orbit and so on $\aeff$
as well.
For the stadium, (c), the principal peak corresponds to the horizontal
orbit, which appears at $4$ because we use only the wavefunctions symmetric
about the vertical symmetry axis (equivalent to using a half-stadium).
Note the excellent agreement between the semiclassical theory and the
numerical results in all cases.
}
\end{center}
\end{figure}

\begin{figure}[t]
\begin{center}
\leavevmode
\epsfxsize = 8.6cm
\epsfbox{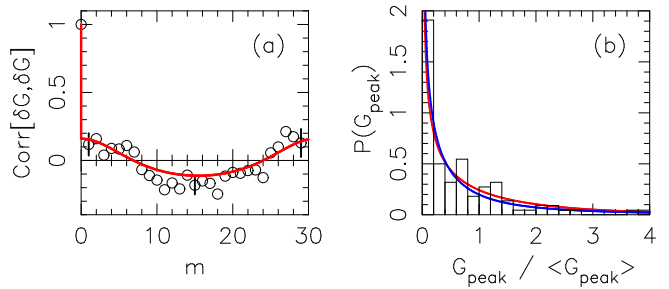}
\end{center}
\caption{
Conductance statistics: (a) peak-to-peak correlation function
and
(b) probability distribution of $G$ for the stadium data in Fig. 1.
The numerical correlation function (circles with typical error
bars)-- the average of all pairs of 
peaks $m$ peaks apart-- is in good agreement with the semiclassical 
theory (red). The agreement for small $m$ is surprising
since this regime is not semiclassical, but shows how dynamics can give
rise to correlations even between 
nearest-neighbors. The numerical
probability distribution (histogram) is for the entire range of data in
Fig. 1 and is compared to both the semiclassical theory (red) and the
standard statistical theory based on random wavefunctions (blue).
The two theories predict nearly the same result for this quantity,
and both are consistent with the numerics.
}
\end{figure}

In Fig. 1 we show an example trace of $G$ for the stadium.
The calculation clearly demonstrates both strong peak-to-peak fluctuations
and an oscillatory modulation of the heights (3 periods are observed). While
the former comes from the quasi-random fluctuations in the wavefunctions
near the leads, the large oscillatory modulation is caused by interference
along the horizontal orbit which connects both leads.

Since the main theoretical result concerns the periodic modulation of the
peak heights, it is natural to consider the Fourier power spectrum of
$G_{\rm peak}(k)$. In Fig. 2 we present a comparison of the numerical and
semiclassical power spectra, calculated for both integrable (circular) and
chaotic (stadium) dots. The data show that for both the circle and the
stadium the power spectrum has sharp peaks corresponding to periodic
orbits. More peaks appear for narrow leads [Fig. 2(a)] because the lack of
momentum constraint in this regime allows coupling to more periodic orbits.
The excellent agreement between the semiclassical expression and the
numerical result in all cases is a striking demonstration of the validity of
our theory.

Further characterization of the peak fluctuations is shown in Fig. 3. The
peak-to-peak correlation function is
\begin{eqnarray} 
{\rm Corr}_m\left[\delta G, \delta G\right]\equiv 
\frac{
\langle 
\delta G\left(E_{n+m}\right) \delta G\left(E_n\right) 
\rangle_n}
{\langle \delta G\left(E_n\right)^2 \rangle_n},
\label{def_correlator}
\end{eqnarray}
where
$
\delta G\left(E_m\right) \equiv G\left(E_m\right) - 
\langle G\left( E_n\right) \rangle_n
$
is a natural measure of the statistics of nearby peaks. A semiclassical
expression for this quantity can be derived by assuming that the
distribution of individual peak heights is locally
Porter-Thomas\cite{ReichlTransition}, with the mean given by the
semiclassical envelope (\ref{gamma_sc}). Indeed, as was first shown by
Kaplan and Heller \cite{KaplanHeller}, this is generally true for
wavefunction fluctuations in chaotic systems. We obtain
\begin{eqnarray}
{\rm Corr}_m & = & 
\delta_{m,0} + \left(1 - \delta_{m,0}\right)\times
\frac{ 
\sum_\mu A_\mu^2
\cos\left( \frac{\tau_\mu \Delta}{\hbar} m \right)}{4 \bar{\Gamma}^2 +
3 \sum_\mu A_\mu^2} \, .
\label{sc_correlator}
\end{eqnarray}
In Fig. 3(a) we compare the semiclassical correlation function with
numerical data for the stadium dot. The oscillatory behavior for large
separations reflects the peak in the corresponding power spectrum in Fig. 2
and is in agreement with the semiclassical result.  The positive correlation
for nearest neighbors is also in agreement with the semiclassical theory,
demonstrating the influence of dynamics even in this apparently
non-semiclassical regime.

When $T \! \gg \! \Delta$, the major source of correlations between
neighboring peaks is the joint contribution of several resonances to the
same conductance peak\cite{Alhassid98}. In this regime the
``nearest-neighbor'' correlator is ${\rm Corr}_{m = 1} \! \sim \! 1$, and
the dynamical effect accounts for only a small correction to the correlation
function. However, for low temperature $T \! \leq \! \Delta$, the
correlations due to temperature are exponentially suppressed. In this
regime, the correlations induced by dynamical modulation dominate, and they
account for the experimentally observed enhancement of correlations at low
temperatures\cite{MarcusPatel98}.

The probability distribution of $G_{\rm peak}$ over a large energy range is
the main quantity considered in the previous statistical
theories\cite{jalabert,nato-book-2}. They predict no peak-to-peak
correlation or periodic modulation of the heights, and a Porter-Thomas
distribution: $P(\Gpeak) = \sqrt{4/\pi \Gpeak} \exp(-\Gpeak)$. Considering
an energy range larger than any period in Eq. (\ref{gamma_sc}), we find, in
contrast, that the distribution should be locally Porter-Thomas but with the
mean modulated by the periodic components, as in Ref.
\onlinecite{KaplanHeller}. Curiously, the resulting distribution is not very
different from Porter-Thomas: Fig. 3(b) shows that the two theories predict
nearly the same result, and both are consistent with numerical calculation.
This explains why no dynamical effect was observed in the experimental
peak-height probability distribution\cite{Chang96,MarcusFolk96}.

In contrast, the periodic modulation of the peak heights has been observed
in several recent
experiments\cite{MarcusFolk96,MarcusCronenwett97,MarcusPatel98}. The
clearest observation is in Ref. \onlinecite{MarcusPatel98}: the
data in their Fig. 1 show modulated peak heights as a
function of the number of electrons in the dot. In their trace of 90 peaks,
approximately six oscillations are visible, yielding a period of 
$\sim \! 15$ peaks. In our treatment, this period is the ratio of the period 
of fundamental oscillation in Eq. (\ref{gamma_sc}) to the level spacing
$\Delta$. The fundamental period is given by
$\left(\frac{1}{\hbar}\frac{\partial S_\mu}{\partial \varepsilon}
\right)^{-1} \equiv \hbar/\tau_\mu$ where $\tau_\mu$ is the period of the
relevant orbit. To determine $\tau_\mu$, we use the billiard approximation:
$L_\mu \equiv v_F \tau_{\mu}$, where $L_\mu$ is the length of the shortest orbit
and $v_F$ is the Fermi velocity. We use the micrograph of the dot to
estimate $L_\mu$ for the $V$-shaped orbit connecting the two leads, and
calculate $v_F$ from the experimental density\cite{Marcus_comm}. Using the
appropriate spin-resolved level spacing $\Delta \smeq 10$ $\mu$eV (which is
half of the spin-full value from excitation measurements in Ref.
\onlinecite{MarcusPatel98}), we find $\hbar/\tau_{\mu}\Delta \! \approx \!
12$. Because the billiard approximation underestimates the period in a soft
wall potential, this is a lower bound for the modulation period, and
therefore our theory is in good agreement with the experiment.

Similarly, we make estimates which are consistent with the other two
experiments showing variation as a function of number of
electrons\cite{Chang96,MarcusFolk96}.  A similar approach to the peak 
modulation as a function
of magnetic field is in agreement with the experimental
results\cite{MarcusFolk96,MarcusCronenwett97} as well.
This agreement with experiment is perhaps surprising, since the adding of
electrons changes the effective potential defining the dot; however,
experiments on "magnetofingerprints" of the peaks\cite{StewartMarcus}
suggest that this change is small while to affect the dynamical modulation
one must substantially change the action of the shortest periodic orbit.

We close with two further experiments suggested by our results. First,
if the tuning parameter used to change the number of electrons, such as
a gate voltage, does not change the action of the dominant periodic orbit, 
then no modulation connected to that orbit should be seen. In particular,
gates which affect different parts of the dot may produce different
oscillatory behavior. Second, several samples made in a robust 
geometry-- a circle with directly opposite leads, for example-- should show 
the same modulation. Any deviations from the same behavior would be
a sensitive indication of the material quality.

We gratefully acknowledge stadium eigenfunction calculations
by J.~H.~Lefebvre, helpful discussions with C. M. Marcus, 
the hospitality of the Aspen Center for Physics where this work was
initiated, and support from the US ONR and the US NSF.

\pagebreak


\begin{references}
\bibitem{subbands} 
The expressions given for $\bar{\Gamma}$ and $A$ are for tunneling from a
single subband in the lead. The generalization to many subbands is
straightforward.
\end{references}
\end{document}